\documentstyle[12pt,draft,nature_macros]{nature}
\font\large=cmbx10 scaled \magstep3

\newcommand{\be}{\begin{equation}}
\newcommand{\ee}{\end{equation}}

\newcommand{\ifm}[1]{\relax\ifmmode#1\else$\mathsurround=0pt #1$\fi}
\newcommand{\kms}{\ifmmode\,{\rm km}\,{\rm s}^{-1}\else km$\,$s$^{-1}$\fi}
\newcommand{\kpc}{\ifmmode\,{\rm kpc}\else kpc\fi}
\newcommand{\Mpc}{\ifmmode\,{\rm Mpc}\else Mpc\fi}

\newcommand{\ltsima}{$\; \buildrel < \over \sim \;$}
\newcommand{\lsim}{\lower.5ex\hbox{\ltsima}}
\newcommand{\gtsima}{$\; \buildrel > \over \sim \;$}
\newcommand{\gsim}{\lower.5ex\hbox{\gtsima}}

\def\no{\noindent}

\def\Mpc3{\,{\rm Mpc}^{-3}}

\def\ifm#1{\relax\ifmmode#1\else$\mathsurround=0pt #1$\fi}
\def\kms{\,{\rm km\,s\ifm{^{-1}}}}

\newcommand{\ad}[1]{}

\begin{document}

\title{At the heart of the  matter: the origin of bulgeless dwarf galaxies and Dark Matter cores}

\author{F. Governato$^1$, C. Brook$^2$,  L.
  Mayer$^{3}$,  A. Brooks$^4$, G. Rhee$^5$, J. Wadsley$^6$, P. Jonsson$^7$, B. Willman$^{8}$,~G. Stinson$^6$,
  T. Quinn$^1$ and P. Madau$^{9}$
 \institute{$^1$Astronomy Department, University of Washington, Seattle, WA 98195, US \\
    $^2$ Peremiah Horrocks Institute, University of Central Lancashire, Preston, Lancashire,PR1 2HE, UK\\
    $^{3}$Institute for Theoretical Physics, University of Zurich,  Winterthurestrasse 190, CH-8057 Z\"urich.\\
    $^4$ Theoretical Astrophysics, California Institute of Technology, MC 350-17, Pasadena, CA, 91125 US\\
    $^{5}$Department of Physics and Astronomy, University of Nevada, Las Vegas, NV US\\
    $^{6}$Department of Physics and Astronomy, McMaster University, Hamilton, ON, L8S 4M1, Canada \\
    $^7$ Institute of Particle Physics, UCSC, Santa Cruz, CA 95064, US \\ 
    $^8$Haverford College, Department of Astronomy 370 Lancaster Ave, Haverford, PA 19041 \\
    $^{9}$Department of Astronomy and Astrophysics. University of California, Santa Cruz, CA  95064 US \\
} }

\dates{}{}
\mainauthor{Governato et al.}
\headertitle{Bulgeless dwarf galaxies and CDM}

\summary{ For almost two decades the properties of ``dwarf'' galaxies
  have challenged the Cold Dark Matter (CDM) paradigm of galaxy
  formation$^{1}$. Most observed dwarf galaxies consists of a rotating
  stellar disc$^{2}$ embedded in a massive DM halo with a near
  constant-density core$^{3}$.  Yet, models based on the CDM scenario
  invariably form galaxies with dense spheroidal stellar ``bulges''
  and steep central DM profiles$^{4,5,6}$, as low angular momentum
  baryons and DM sink to the center of galaxies through accretion and
  repeated mergers$^7$.  Processes that decrease the central density
  of CDM halos$^8$ have been identified, but have not yet reconciled
  theory with observations of present day dwarfs. This failure is
  potentially catastrophic for the CDM model, possibly requiring a
  different DM particle candidate$^{9}$.  This Letter presents new
  hydrodynamical simulations in a $\Lambda$CDM framework$^{10}$ where
  analogues of dwarf galaxies, bulgeless and with a shallow central DM
  profile, are formed.  This is achieved by resolving the
  inhomogeneous interstellar medium, resulting in strong outflows from
  supernovae explosions which remove low angular momentum gas. This
  inhibits the formation of bulges and decreases the dark-matter
  density to less than half within the central kiloparsec. Realistic
  dwarf galaxies are thus shown to be a natural outcome of galaxy
  formation in the CDM scenario.  }

\maketitle

\noindent

\newpage

In $\Lambda$CDM, the favored theory of cosmic structure formation,
galaxy discs form as gas cools and collapses inside spinning halos of
collisionless DM, reaching centrifugal equilibrium and turning into
stars$^{11}$.  Models which assume that the stellar component of
galaxies inherits the angular momentum distribution of their host DM
halos also predict the formation of a centrally concentrated stellar
bulge and a cuspy DM profile$^{7,12}$. In contrast, the vast majority of dwarf
galaxies have no stellar bulges, and the observed rotation curves of
small galaxies often rise almost linearly in the central kpc, a result
interpreted as a sign of a shallow DM distribution$^{13,14}$.  This
excess of low angular momentum material creates the so called "angular
momentum problem"$^{15}$ for CDM models.

A proposed solution to the existence of bulgeless galaxies invokes gas
winds created by multiple supernovae (SN) explosions to selectively
remove low angular momentum baryons from the center of
galaxies$^{16}$.  SN winds are observed in both local and high
redshift galaxies and are efficient at removing gas from the discs of
nearby galaxies at a rate of few times the current star formation (SF)
rate$^{17,18}$.  Modeling the formation of a highly inhomogeneous
multi phase ISM is necessary to tie SF to high density gas regions and
to create SN winds able to affect the internal mass distribution of
galaxies$^{19,20}$.  Such numerical schemes for SF and resulting
feedback have been applied to the formation of high redshift
protogalaxies, leading to significant baryon loss and less
concentrated systems$^{8,20}$.  Similarly, dynamical
arguments$^{21,22}$ suggest that bulk gas motions (possibly SN
induced) and gas clouds orbital energy loss due to dynamical friction
can transfer energy to the center of the DM component. Sudden gas
removal through outflows then causes the DM distribution to expand.
These mechanisms were demonstrated to operate effectively in small
high-redshift halos of total mass around 10$^9$ M${\odot}$, where they
create small DM cores$^{8}$.  However, such methods and the required
high resolution have not been applied to cosmological hydrodynamical
simulations of present day dwarf galaxy systems (V$_{rot}$ $\sim$ 60
$\kms$).  Showing that the properties of dwarf galaxies can be
accurately predicted by the CDM scenario would end the ``small scale
crisis'' and further constrain the properties of the DM particle
candidate.

To study the formation of dwarf galaxies in a $\Lambda$CDM cosmology.
we analyze a novel set of cosmological simulations. Baryonic processes
are included, as gas cooling$^{8}$, heating from the cosmic UV field$^{23}$,
star formation and SN driven gas heating (Figure 1). Resolution is
such that dense gas clumps as small as 10$^5$ M$_{\odot}$ are
resolved, similar to real SF regions$^{19}$. Hence stars are allowed
to form only in cold gas regions with a local density higher than 100
amu/cm$^{3}$, reflecting typical conditions in real galaxies.  This
description of SF is a critical improvement over many previous
cosmological simulations, where star forming regions were not
individually resolved. A single dwarf galaxy (DG1) is analyzed, but we
obtained equivalent results with galaxies of similar mass and
different assembly histories and halo spin (another simulated galaxy
is described in the supplemental material).  DG1 has a rich merger
history: three proto-galaxies of similar mass merge at z$\sim$3, and a
large satellite is accreted at z$\sim 1.2$ that has a mass 1/3 that of
the central galaxy (a ``major'' merger). Several other satellites are
accreted, including one at low redshift.  Star formation is bursty, as
observed in nearby dwarfs$^{27}$ peaking at 0.25 M$_{\odot}$/yr during
interactions at z $\sim$ 2. The disc component assembles shortly after
that (Figure 1, top). At z=0 an exponential stellar disc is surrounded
by an HI disk which extends out to six disc scale lengths. The galaxy
shows no sign of a stellar spheroid (Figure 1, bottom right).  The
star formation rate declines after z=1, and by the present time it is
down to 0.01 M$_{\odot}$/yr, in agreement with galaxies of similar
magnitude.

SN feedback creates holes in the HI distribution due to bubbles of hot
gas expanding perpendicular to the disc with velocities approaching
100 $\kms$ (Figure 1, top left).  The HI super shells close to the
disc plane are typically a few hundred parsecs wide, expanding at
10-30 $\kms$, similar to those observed in dwarfs$^{28}$.  As SF
happens in short, spatially concentrated bursts including several
coeval star particles, the typical energy per unit mass released in
the surrounding gas is sufficient to disrupt gas clouds and generate
gas fountains that unbind gas from the shallow potential of the galaxy
at 2-6 times the instantaneous SF rate, consistent with
observations$^{17}$. As predicted in earlier studies$^{20}$ feedback
from spatially resolved SF results in a realistic low SF efficiency
and a total baryon mass loss equal to a few times the final amount of
stars. As star forming regions are centrally biased or rapidly sinking
to the galaxy center due to dynamical friction, most of the gas
becoming unbound is preferentially removed at small radii and at
z$>$1.

Mock images$^{29}$ (Figure 1, lower panels) show that in redder bands
the optical disc is relatively featureless, although star forming
regions are visually associated with short lived spiral arms.  The
striking feature of this galaxy is the complete absence of a stellar
spheroid even when observed edge-on (Figure~1).  The radial light
distribution in all optical and near-IR bands has an almost perfect
exponential profile as observed in dwarf galaxies.  This galaxy would
thus be classified as ``bulgeless''$^2$, i.e. lacking of a visible
central stellar spheroid.  The formation of a pure disc galaxy with
structural properties typical of observed gas rich dwarfs$^{24}$ is a
fundamental success of this set of simulations.

The underlying DM and baryonic mass profile of DG1 has been measured using
kinematic estimators. The rotation curve of DG1 (Figure 3), was
obtained measuring the rotational motion of cold (T$<$ 10$^4$ K) gas
as a function of radius using the ``tilted ring analysis'', which
reproduces the effects of observational biases such as disc
distortions and warping, bars and pressure support from non circular
motions$^{30}$. The rotation curve of DG1
rises almost linearly out to one stellar disc scale length, and is
still rising at four scale lengths ($\sim$ 4 kpc), similar to the
rotation curves of real dwarf galaxies. This is a great success as
simulations have persistently produced rotation curves which rise
rapidly in the inner regions and peak inside one scale length,
symptomatic of their steep central mass distributions$^{5}$.
The DM central density of DG1 has a shallow profile over a
``core'' of roughly 1 kpc in size (see figure 3 and Figure 5 in
Supplementary Info), comparable to those measured in many dwarf
galaxies$^{3,,13,14}$.  Accordingly, the DM density averaged over the same
radius is 10$^{7.5}$ M$_{\odot}$kpc$^{-3}$, about 50\% lower than in a
control run where the gas is not allowed to cool or form stars, and
where the slope of the inner DM profile is instead steep as in typical
DM only simulations$^{6}$.

Outflows are the main mechanism in altering the central density
profile of the baryonic component of galaxy DG1. The strongest
outflows correlate with SF bursts, caused by mergers and strong
interactions, when dense gas regions form from disk instabilities and
sink to the center due to dynamical friction.  SN feedback destroys
these gas clumps as soon as they start forming stars. Outflows then
selectively remove most low angular momentum gas before it is
transformed into stars, effectively quenching the processes that would
lead to a concentrated baryon distribution and to the formation of
stellar bulges (see Supplemental material). By the present time the
angular momentum distribution of stars formed from the remaining gas
has a median value higher than the DM, and lacks its low angular
momentum tail (Figure 4).

The removal of centrally concentrated, low angular momentum gas is
also closely connected to the origin of shallow DM profiles.  As the
galaxy DG1 assembles, gas starts collecting at its center via clumps
and filaments while the DM remains smoothly distributed.  This spatial
decoupling between the gas and DM can lead to efficient orbital energy
transfer from the gas to the DM through gas bulk motions$^{8}$ and gas
orbital energy loss.  Additionally, the gas outflows ensuing from
subsequent SF rapidly remove a large fraction of the gas, leading to a
significant loss of DM binding energy, causing a net expansion and the
formation of a shallow DM profile$^{21,22}$. Our simulations provide
direct confirmation of these two mechanisms, as the expansion of
the collisionless DM component occurs over several Gyrs, closely
following the strongest outflows.  Outflows happen both in smaller
mass progenitors and then at the center of the main galaxy, where the
process of core formation is essentially complete by z $\sim$ 0.5 when
the DM profile in the inner $\kpc$ settles to a shallow slope with
$\rho$ $\propto$ r$^{-0.6}$, comparable to those observed and
shallower than a DM--only control run that has the canonical, much
steeper profile (see Figure 3 and Supplemental info).

These results predicts that low-mass, bulgeless disc galaxies should
also have a shallow DM central profile and predicts that these two
properties should be correlated in observed samples of nearby
galaxies, as they are originated by the same physical processes.

\begin{figure}
\vskip 11.6cm 
{\includegraphics{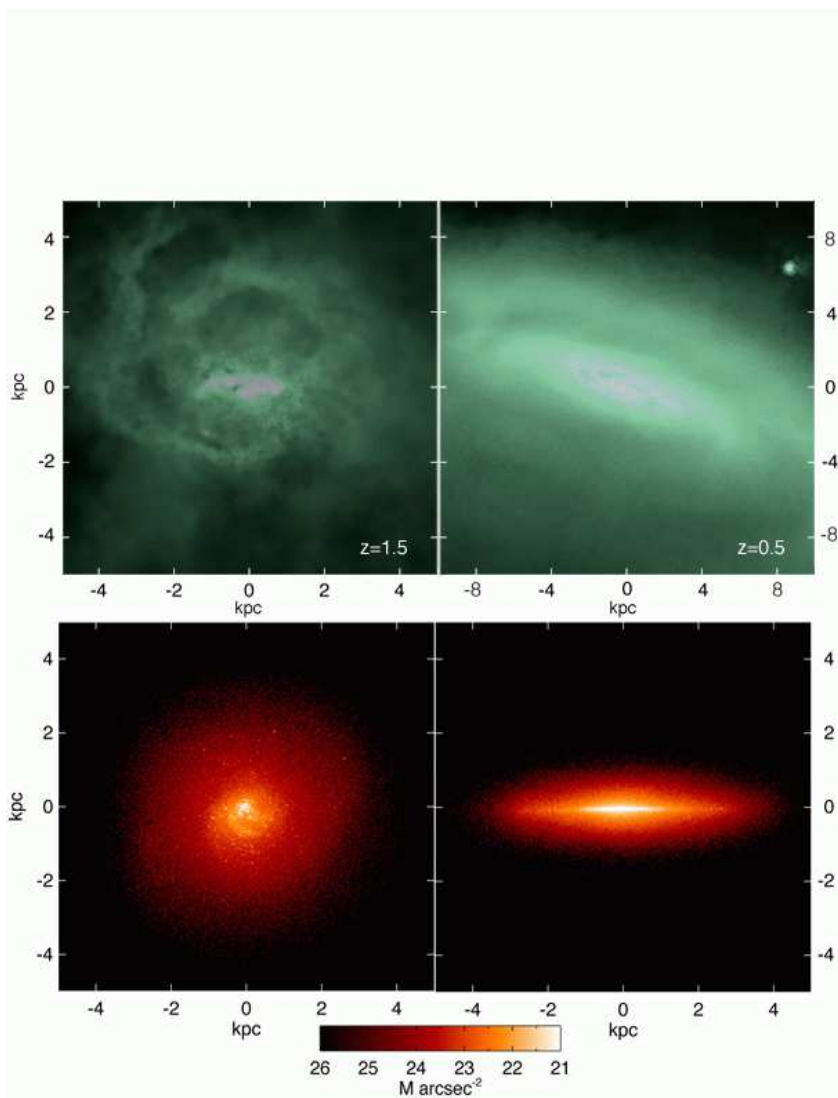}}
\caption[]{\small The observable properties of simulated galaxy DG1.
  Top Left: Colour-coded density map of the gas distribution at
  z$=$1.5, showing the gas outflows and super shells. Top Right: The
  gas distribution at z$=$0.5 when the disc has fully formed (note the
  larger scale).  At z$=$0 the total mass of the system within the
  virial radius is 3.5 $\times$ 10$^{10}$ M$_{\odot}$. As a result of
  outflows and inefficient star formation, the disk (including HI gas
  and stars) to virial mass ratio is only f$_{disk}$= 0.04,  70\%
  of the disk mass  is in HI, and the M$_{HI}$/L$_B$ ratio is 1.2.  The
  amount of baryons within the virial radius  is only
  30\% of the cosmic fraction. These values are consistent with those
  observed in real galaxies of similar mass$^7$.  Bottom Left: The
  face on light distribution at z$=$0 in the SDSS $i$ band.  Bottom
  Right: The galaxy seen edge on in the same band. The effect of dust
  absorption is included. The total magnitude of the galaxy in the $i$
  SDSS band is -16.8 giving an $i$ band M/L ratio of $\sim$ 20. The
  galaxy $g-r$ colour is 0.52, typical of star forming dwarf
  galaxies$^{24}$. The rotation velocity is $\sim$ 55 $\kms$, as
  measured using the W$_{20}$/2 line width of the HI distribution.
  This simulation resolves the internal structure of galaxy DG1 with
  several million resolution elements, achieves a mass resolution of
  10$^3$ M$_{\odot}$ for each star particle and a force resolution of
  86pc.  In the supplementary material we show that high resolution
  coupled with SF being spatially associated to small gas clouds is a
  fundamental requirement for SN feedback to originate outflows and
  lower the density at the center of galaxy halos.  Simulations using
  the same implementation of SF and feedback reproduce some global
  scaling properties of observed galaxies across a range of masses and
  redshifts$^{25,26}$.  }
\label{fig:1}
\end{figure}

\begin{figure}
\vskip 14.6cm 
{\includegraphics{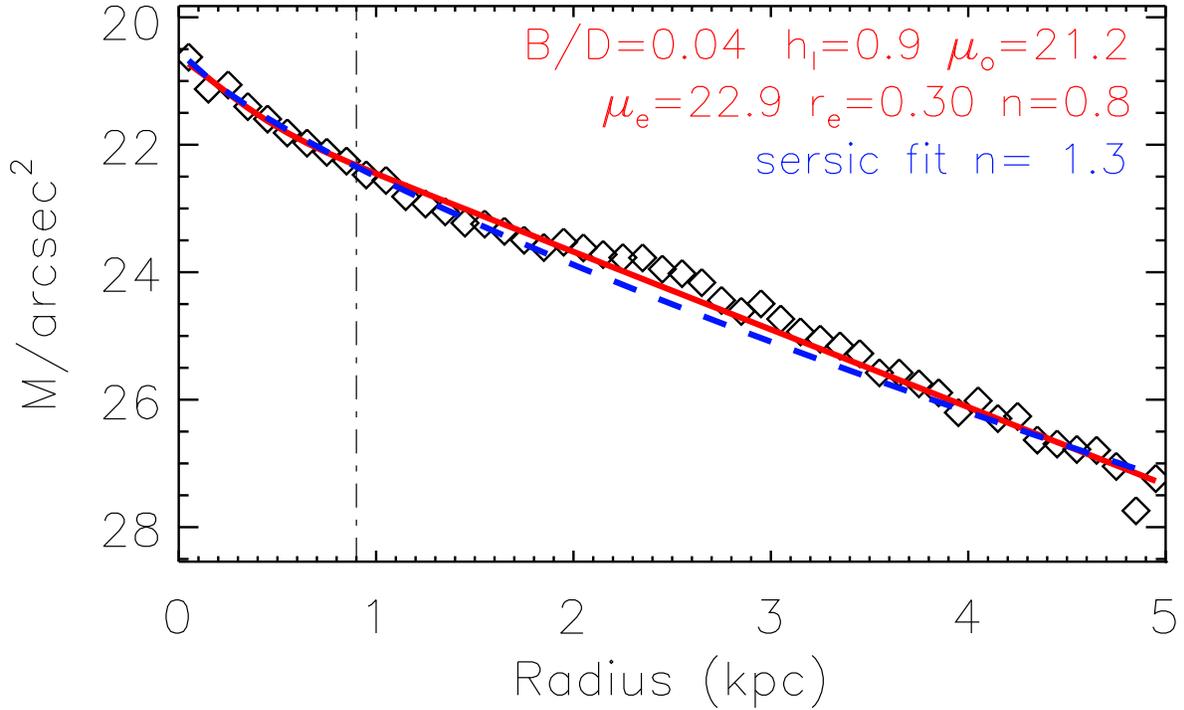}}
\caption[]{\small The SDSS {\it i}-band radial light profile of the
  simulated dwarf galaxy DG1 at z$=$0.  Being less sensitive to recent
  star formation events, the $i$ band is commonly used to describe the
  underlying stellar mass distribution. The galaxy has an almost pure
  exponential disc--like light profile (diamonds), with a formal
  bulge+disc fit to the 2D light distribution giving a B/D ratio of
  0.04 as seen face on.  The measured disc scale length is 0.9 kpc
  (dot dashed vertical line). The dashed and continuous lines show the
  Sersic+exponential fit to the disc and a one component Sersic
  profile fit with n$=$1.3. An index value of n$<$1.5 identifies
  bulgeless disc galaxies in large surveys$^{2,3}$. The galaxy would
  then be classified as ``bulgeless''. Images in other optical bands
  and the near--IR K band give a similar fit with extremely low B/D
  ratios.  Fits were measured using the public software GALFIT and a
  1D two-component fitting procedure, obtaining similar results. A
  second galaxy (DG2), shows a profile best fit by a pure exponential,
  with a Sersic index of one (see Figure 6 in Supplementary info and
  references there).  Images were created using {\sc SUNRISE}$^{29}$,
  which creates spectral energy distributions using the ages and
  metallicities of each simulated star particle, and takes into
  account the full 3D effects of dust re-processing.}
\label{fig:2}
\end{figure}

\begin{figure}
\vskip 10.6cm 
{\includegraphics{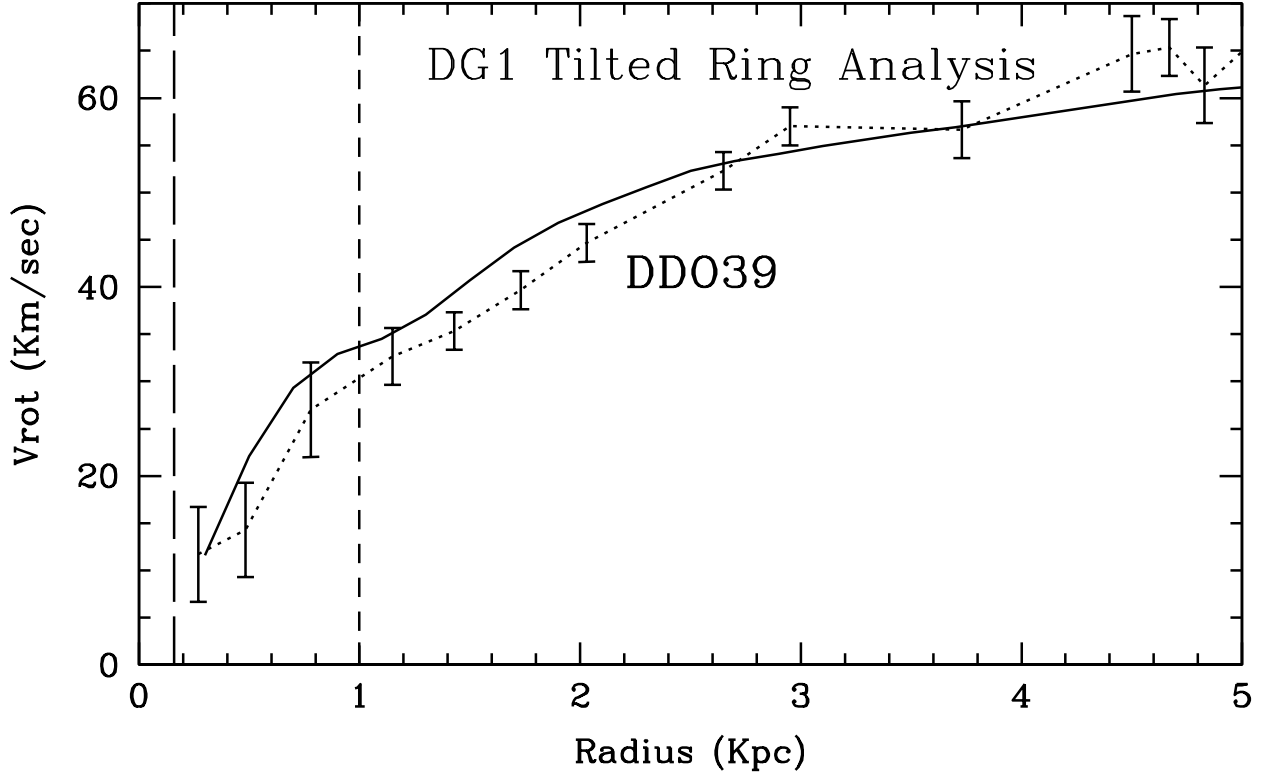}}
\caption[]{\small The rotation curve of the simulated dwarf compared
  to that measured for a real galaxy.  The continuous black line shows
  the rotation velocity of the galaxy using actual projected velocity
  field and the tilted ring analysis$^{30}$ (see Supplemental Info).
  The dotted line shows the rotation curve of the galaxy DDO39 as
  measured using a similar technique$^{14}$, with standard deviation
  error bars. The velocity profile of both the observed and simulated
  galaxies imply a DM distribution with a core scale length of about 1
  $\kpc$, as directly measured in the simulation (Fig.5 in the
  Supplemental Info). The long dashed vertical line shows the force
  resolution of the simulation, while the dashed vertical line marks
  the approximate scale length of the DM ``core''. The underlying DM
  density $\rho_{DM} \propto r^{\alpha}$, with $\alpha$=-0.6 in the
  central kpc, is consistent with observational estimates and
  shallower than a DM only simulation (see Supplemental info) that
  would predict a steeper profile with $\alpha = -1.3$ }
\label{fig:3}
\end{figure}

\begin{figure}
\vskip 11.6cm 
{\includegraphics{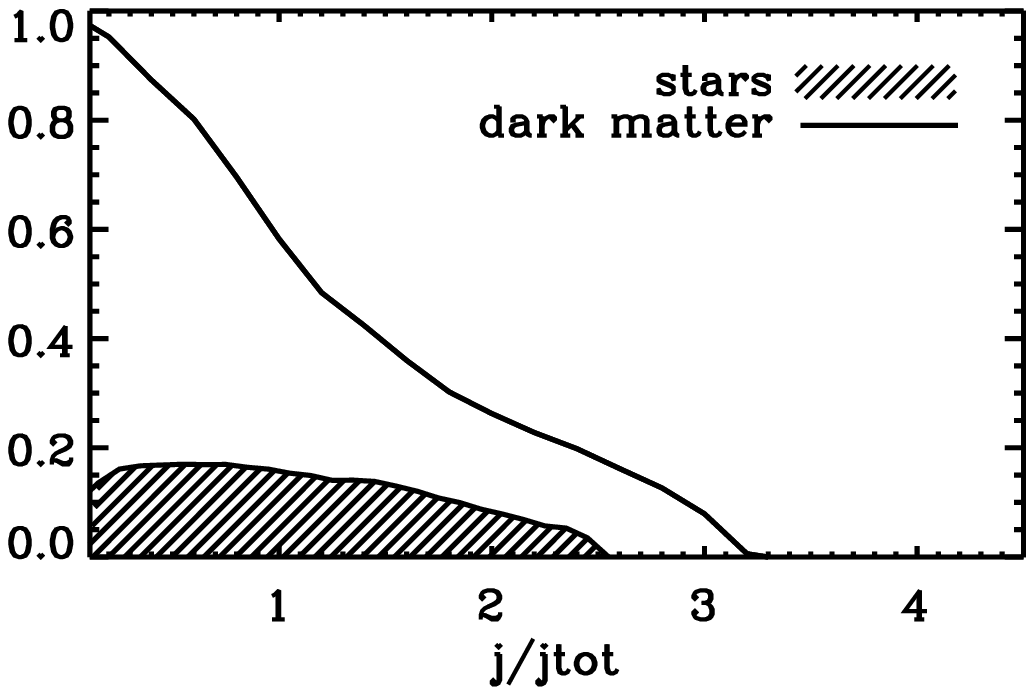}}
\caption[]{\small A comparison between the angular momentum
  distribution of the stellar disk and the dark matter halo in the
  simulated galaxy DG1.  The panel shows the present day time angular
  momentum distribution of disc stars (shaded area) and DM particles
  (area below the continuous line) normalized by the median value for
  the whole DM halo (jtot$=$1). The angular momentum distribution of
  star particles has a narrower distribution, a higher average and
  significantly less low angular momentum material than the DM, due to
  centrally concentrated outflows preferentially removing low angular
  momentum baryons.  As a result, the radial stellar distribution is
  similar to that measured for normal dwarf galaxies$^{2,3}$ (see also
  Figure 2).}
\label{fig:3}
\end{figure}

\addtolength{\baselineskip}{-0.05\baselineskip}


\newpage

\noindent{\bf Supplementary Information} is linked to the online version
of the paper at www.nature.com/nature

\noindent{\bf Acknowledgments} We acknowledge discussions with
L. Blitz, A. Kravtsov, J. Primack, I. Trujillo and V. Wild. We thank
R. Swaters and the THINGS team for sharing some of their data with
us. LM was supported by a grant of the Swiss National Science
Foundation. LM and CB thank the Kavli Institute for Theoretical
Physics at UC Santa Barbara for hospitality during the early stages of this
work. FG and PM thank the computer support people at Nasa Advanced
Supercomputing, TERAGRID, ARSC, and UW, where the simulations were
run.


\noindent {\bf Author Information} Reprints and permissions
information is available at\\
 npg.nature.com/~reprintsandpermissions. The
authors declare no competing financial interests.

Correspondence and requests for materials should be
addressed to F.G. (fabio@astro.washington.edu).


\vfill\eject
\centerline{\bf SUPPLEMENTARY INFORMATION}


\bigskip\no This is an extension of the Letter to Nature, aimed at
providing further details, in support of the results reported in its
main body.

\section{The SPH Treecode Gasoline}

We have used the fully parallel, N-body+smoothed particle
hydrodynamics (SPH) code GASOLINE to compute the evolution of both the
collisionless and dissipative component in the simulations. A detailed
description of the code is available in the literature$^{31}$. Here we
recall its essential features.  GASOLINE computes gravitational forces
using a tree--code$^{32}$ that employs multipole expansions to
approximate the gravitational acceleration on each particle. A tree is
built with each node storing its multipole moments.  Time integration
is carried out using the leapfrog method, which is a second-order
symplectic integrator requiring only one costly force evaluation per
timestep and only one copy of the physical state of the system.  In
cosmological simulations periodic boundary conditions are
mandatory. GASOLINE uses a generalized Ewald method$^{33}$ to arbitrary
order, implemented through hexadecapole.

SPH is a technique of using particles to integrate fluid elements
representing gas$^{34,35}$.  GASOLINE is fully Lagrangian, spatially and
temporally adaptive and efficient for large $N$. The version of the
code used in this Letter includes radiative cooling and accounts for
the effect of a uniform background radiation field on the ionization
and excitation state of the gas. The cosmic ultraviolet background is
implemented using the Haardt-Madau model$^{23}$, including
photoionizing and photoheating rates produced by PopIII stars, QSOs
and galaxies starting at $z=9$. The assumption that reionization
occurred at high z is consistent with the combination of the 3rd year
WMAP results and the Gunn-Peterson effect in the spectra of distant
quasars$^{36}$. We use a standard cooling function for a primordial
mixture of atomic hydrogen and helium at high gas temperatures, but we
include the effect of metal cooling$^{8}$ and evolving gas
metallicities below $10^4$ K (the metallicity in dwarfs is indeed much
lower than solar$^{37}$, with $-1 < [Fe/H] < -2$).  The internal energy
of the gas is integrated using the asymmetric formulation, that gives
results comparable to the entropy conserving formulation$^{38}$ but
conserves energy better. Dissipation in shocks is modeled using the
quadratic term of the standard Monaghan artificial viscosity$^{35}$.  The
Balsara correction term is used to reduce unwanted shear
viscosity$^{39}$.

In the simulations described in this Letter star formation occurs when
cold gas reaches a given threshold density$^{40,41}$ typical of actual
star forming regions (we used 100 amu/cm$^3$ in most runs and discuss
the effect of a lower threshold in \S 5). SF then proceeds at a rate
proportional to $\rho_{gas}^{1.5}$, i.e. locally enforcing a Schmidt
law. The adopted feedback scheme is implemented by releasing energy
from SN into the gas surrounding each star particle$^{41}$. The energy
release rate is tied to the time of formation of each particle (which
effectively ages as a single stellar population with a Kroupa IMF).
To model the effect of feedback at unresolved scales, the affected gas
has its cooling shut off for a time scale proportional to the Sedov
solution of the blastwave equation, which is set by the local density
and temperature of the gas and the amount of energy involved. In the
high resolution runs described in this study this translates into
regions of $\sim$ 0.1 to 0.3 kpc in radius being heated by SN feedback
and having their cooling shut off. At z $>$1 the affected gas has its
cooling shut off for typically 5-10 million years. However, even
during high z starbursts only a few per cent of the gas in the disc
plane has a temperature T $>$ 40,000 Kelvin.  The effect of feedback
is to regulate star formation in the discs of massive galaxies and to
greatly lower the star formation efficiency in galaxies with peak
circular velocity in the 50 $<$ V$_c$ $<$150 $\kms$ range$^{25}$. At
even smaller halo masses (V$_c$ $<$ 20-40 $\kms$) the collapse of
baryons is largely suppressed by the cosmic UV field$^{25,42}$. Other
than the density threshold only two other parameters are needed, the
star formation efficiency ($\epsilon$SF = 0.1) and the fraction of SN
energy coupled to the ISM (eSN=0.4).  Similar values have been used in
previous works by this group$^{4}$. However, here we slightly
increased $\epsilon$SF from 0.05 to 0.1 to ensure a better
normalization to the radial Schmidt law.  We
explored values of eSN as high as 0.6 and cooling shutoff times
changing by a factor of a few, verifying that, in the mass range of
this study, mass profiles and SF histories are relatively insensitive
to longer (shorter) cooling shutoff timescales if a smaller (larger)
fraction of SN energy is coupled to the ISM.

\section{Initial Conditions: Cosmological volume and the adopted
  cosmology}

At z $=$ 0 the virial mass of the halos that we studied in this Letter
are 3.5 (DG1) and 2.0 (DG2) $\times$ 10$^{10}$ M$_{\odot}$ (the virial
mass is measured within the virial radius R$_{\rm vir}$, the radius
enclosing an overdensity of 100 times the cosmological critical
density). The halos were selected within a large scale, low
resolution, dark matter only simulation run in a concordance, flat,
$\Lambda$-dominated cosmology: $\Omega_0=0.24$, $\Lambda$=0.76,
$h=0.73$, $\sigma_8=0.77$, and $\Omega_{b}=0.042^{10,43}$.  The size of
the box, 25~Mpc, is large enough to provide realistic torques for the
small galaxies used in this work.  The power spectra to model the
initial linear density field were calculated using the CMBFAST code to
generate transfer functions.

\section{Volume renormalization technique}

The simulations described here require a large dynamical range, from
sub-kpc scales to correctly describe the internal dynamics of the
dwarf galaxies, to several tens of Mpc to include the effects of
cosmic torques from the large scale structure. This is achieved by the
volume renormalization (or ``zoom in'') technique$^{44}$.  From the
z$=$0 output of the large scale simulation the regions of interest
where each galaxy forms were identified and traced back to a
Lagrangian sub region in the initial conditions.  Then the initial
conditions are reconstructed using the same low-frequency waves
present in the low resolution simulation but adding the higher spatial
frequencies. The power spectrum is realized by using Fast Fourier
Transforms (FFTs) to determine displacements of particles from this
grid.  We use a particular technique that allows one to calculate high
resolution FFTs only in the regions of the simulations were mass and
force resolution need to be high. To reduce the number of particles
and make the full N-Body + SPH simulation possible with the same
cosmological context, we construct initial conditions where the mass
distribution is sampled at higher resolution (less massive particles
on a finer grid) and then more coarsely as the distance from the
chosen object increases.  Dark matter particle masses in the high
resolution regions are 1.6$\times$10$^4$ M$_{\odot}$, and the force
resolution, i.e., the gravitational softening, is 86 pc (more details
in Table 1). In total, at z$=$0 there are $3.3 \times 10^6$ particles
within the virial radius of galaxy DG1.  For all particle species, the
gravitational spline softening, $\epsilon(z)$, was evolved in a
comoving manner from the starting redshift (z $\sim$ 100) until z=8,
and then remained fixed at its final value from z=8 to the present.
The softening values chosen are a good compromise between reducing two
body relaxation and ensuring that disc scale lengths and the central
part of dark matter halos will be spatially resolved. Integration
parameter values were chosen according to the results of previous
systematic parameter studies$^{45}$.  This technique achieves CPU
savings of several orders of magnitude, since running the original
volume at the same resolution of the central region would have
required more than 30 billion particles.  This method has been
successfully used in a wide range of cosmological studies$^{4,25,46}$
that have shown how the assembly history of the ``zoomed-in'' objects
is preserved when compared to the original version in the low
resolution volume.

 
\section{The Photometric and Kinematic Analysis}

To properly compare the outputs from the simulation to real galaxies
and make accurate estimates of the {\it observable} properties of
galaxies, we used the Monte Carlo radiation transfer code
{\it{SUNRISE}}$^{29,47}$ to generate artificial optical images (see
Figure~1 and 6) and spectral energy distributions (SEDs) of the outputs of
our run.  {\it{SUNRISE}} allows us to measure the dust reprocessed SED
of every resolution element of the simulated galaxies, from the far UV
to the far IR, with a fully 3D treatment of radiative
transfer. Filters mimicking those of the SDSS survey$^{48}$ are used
to create mock observations. The 2D light profiles are then fitted as
the superposition of a Sersic and an exponential component using the
widely used program GALFIT$^{49}$. Results were compared with a two
component, radially averaged 1D routine developed by the authors,
finding similar results.  

The model galaxy rotation curves (Figure~3) were
determined using the standard tilted-ring analysis$^{50}$ that is
applied to galaxy observations. The idea behind this is to facilitate a
comparison between galaxy models and observations by deriving rotation
curves for both using the same methods. The assumption is that a
rotating disc galaxy can be described by a set of concentric
rings. Each ring has a constant circular velocity and two orientation
angles. The values of the ring parameters are determined from the
observed radial velocities in a set of concentric elliptic annuli. For
each ring we determine the intensity weighted velocity along the line
of sight at each pixel on the annulus and compute the rotational
velocity of the gas at each annulus. The method has been recently
applied to model galaxies$^{51}$ to study the effect of systematic
biases due to non circular motions in the measured rotation curves of
real galaxies$^{52}$.

\section{Resolution tests and cosmic variance}

Recent work has highlighted the role of numerical resolution and
different SF and feedback implementations in driving the structural
properties of simulated galaxies and specifically their internal mass
distribution$^{53,54,55}$.  To demonstrate the robustness of our
results analyse one of our galaxies (DG1) at lower resolutions, and a
different galaxy with a different merger history (DG2), but similar
mass and halo spin ($\lambda$ $\sim$ 0.05$^{56}$), and show that our key results
are  produced in these simulations a) independently of the
assembly history of the parent halo, b) when the spatial and force
resolution are sufficiently high and c) when SF is spatially resolved,
meaning that it happens only in cold gas particles dense enough to be
representative of star forming regions, i.e. with density higher than
100 atom/cm$^3$.

The parameters of the analysed galaxies are summarized in Table~1,
where the galaxy described in the main section of the Nature Letter is
DG1, its lower resolution versions using the same initial conditions
are DG1MR and DG1LR, and a version using a lower density threshold
(0.1 amu/cm$^3$) for star formation is DG1LT (it has the same
resolution as DG1MR).  DG1DM is a run using the same initial
conditions but including only the dark matter component.  DG2 has the
same mass and spatial resolution as DG1. Each run was completed
following the guidelines described in the Initial Conditions
section. The structural properties of the galaxies are measured using
techniques that mimic observations and are presented in Table~2.

As shown in Figure~5, poor resolution and an inconsistent
implementation of star formation affect the global mass distribution
of galaxies, leading to overly dense central regions.  The rotational
velocities based on the local 3D potential (and assuming circular
orbits) for different DG1 realizations and for DG2 are plotted in the
left panel of Figure~5. Convergence is apparent in DG1 and DG1MR
(particles in the DG1MR run are 2.3 times more massive). DG2, which
has a final mass slightly smaller than DG1, also has a velocity
profile that keeps raising for several kpcs.  As a further test, we
verified that the DM only run (DG1DM) has a DM profile similar to what
obtained in previous, similar simulations$^{57}$, with a central DM
density proportional to r$^{\alpha}$, with $\alpha$ $\sim$ -1.3 The
above result shows that in our high resolution simulations the mass
distribution of the galaxies presented in this Letter is free from
numerical effects.

Instead, the low resolution run DG1LR forms a galaxy much more
centrally concentrated. This run adopts the {\it same} SF criteria and
feedback scheme as the higher resolution counterparts, but uses 64
times less particles than DG1 (force resolution is also 4 times worse,
350pc). DG1LR has a flat, rather than rising rotation curve. As we
have shown in previous works, its low resolution causes enhanced
angular momentum loss in the baryonic component via a variety of
numerical effects$^{53}$.  These numerical effects come from the low
number of resolution elements and exist irrespective of the modeling
of the ISM and of the threshold density used for star
formation. Moreover, at the resolution of DG1LR, a single gas particle
has the mass of a SF region (10$^{5-6}$M$_{\odot}$). As individual
density peaks are unresolved, SF only happens at the very center of
the galaxy, further increasing the final central density of baryons
and DM.

Similarly, to show how resolution, details of SF and outflows specifically affect
the DM density profile, the various realizations of DG1 and DG2 are shown
in the right panel of Figure~5. DG1, the main simulation described in
the Nature Letter is in blue, while the medium resolution simulation
is in red, again showing shallow central DM profiles and clear
convergence at our maximum resolution. DG2 has a shallow profile as
well. On the contrary, the dot dashed black line shows the steep
central profile of DG1DM, the DG1 dark matter only simulation, and
highlights the degree to which the DM profile is flattened by
processes connected to the baryonic component.

In our high resolution simulations, where T$_{min}$ $\sim$ 500K we
resolve the local Jeans length implied by the high density threshold
with at least 2 SPH kernels, thus preventing artificial
fragmentation$^{58}$ as visually evident in the top right panel of
Figure 1, showing a smooth gaseous disc. In DG1LR the resolution would
not be enough to resolve the local Jeans length at the high density
threshold, but the large gravitational softening (larger than both the
smoothing length and the Jeans length) has a dominant effect,
suppressing artificial fragmentation, Hence none of our runs suffers
from artificial fragmentation and any clumpiness present is to be
regarded as physical.

It is important to highlight that the the mass distribution of galaxy
DG2 is similar to DG1, as the rotation curve is linearly rising for
several scale lengths and the DM profile shows a shallow central
region of about one kpc in radius.  In panels~a and b of Figure~6, we
plot the face and edge on surface brightness maps of DG2, and the
resulting {\it i}-band surface density profile in panel~c.  Clearly,
the profile is a pure exponential all the way to the center. A two
component fit gives B/D=0.01, and a fit with a Sersic profile yields
an index of 1, corresponding to a pure exponential disc.  At
difference with DG1, DG2 has a quiet merging history with fewer major
mergers and only very minor accretion events after z$\sim$ 1.5,
showing that the suppression of bulge formation and a flat central DM
profile depend more on the strength of feedback rather then on
cosmic variance and the details of the recent assembly history of a
given galaxy.

\begin{figure}
\vskip 11.cm
{\includegraphics{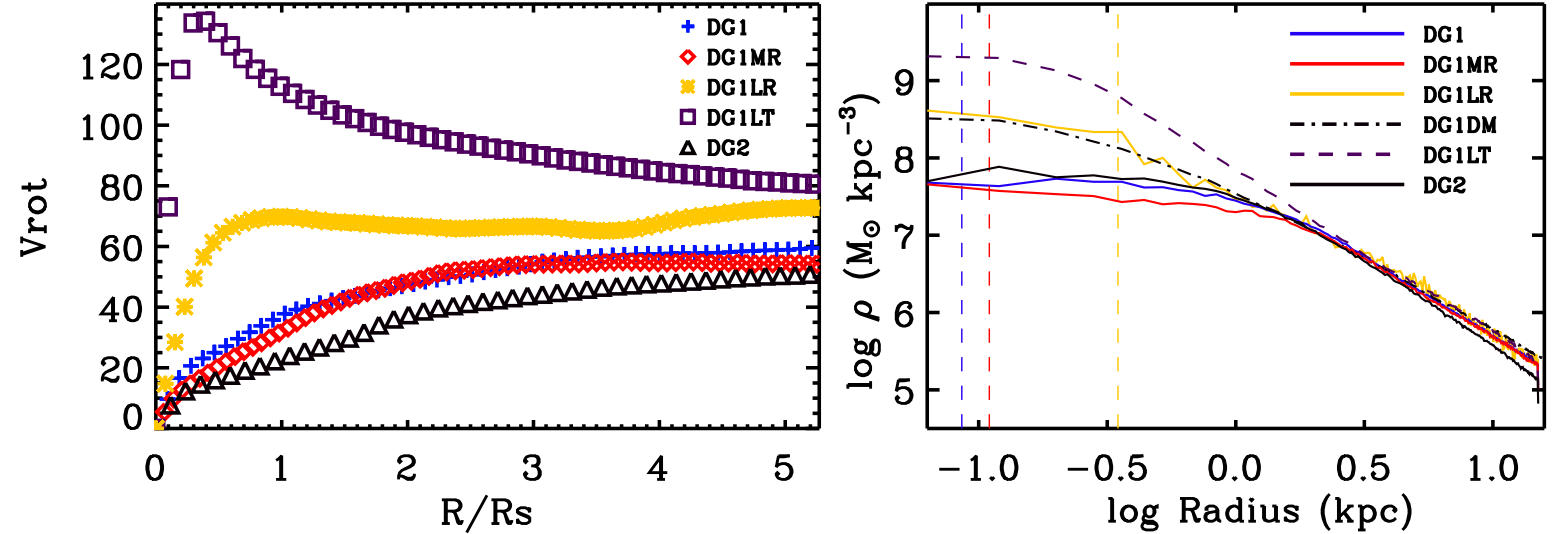}}
\caption[]{\small The rotation curve and DM radial distribution of the
  models described in \S3 and \S5, to show the effects of resolution
  and different SF recipes on the central mass distribution of
  simulated dwarf galaxies. The left panel shows the rotation curve,
  derived using the 3D potential of each galaxy and measured at z$=$0,
  for DG1 (blue crosses), DG1MR (red diamonds) and DG1LR, (yellow
  stars), plotted versus the disc scale length (1kpc for DG1, 0.5kpc
  for DG2). The three runs use the same gas density threshold for SF
  (100 atoms/cm$^3$), but MR and LR runs use only 40\% and 12.5\% of
  the particles of the reference run (with particle masses rescaled to
  the same total mass), and a softening respectively 1.33 and 4 times
  larger. Results have converged at the DG1MR resolution, while the LR
  run shows an excess of central material that is due to poor
  resolution causing artificial angular momentum loss$^{53}$. The
   squares show DG1LT, where star formation is allowed in
  regions with a much lower local density (0.1 atoms/cm$^3$), again
  resulting in a much higher central mass density, due to the lack of
  outflows. This result demonstrates that the correct modeling of
  where SF is allowed to happen (namely only in gas with density
  comparable to that of real star forming regions) is crucial to
  obtain the results described in this Letter. The rotation curve of
  galaxy DG2 (black triangles) shows the same shape as that of the DG1
  run.  Panel B uses a similar colour scheme and plots the DM density
  profile for the same runs. DG1 (blue solid), DG1MR (red dashed) show similar
  profiles, with a DM core of about 1kpc. Color coded vertical lines
  mark the force resolution for each run. The DM only run DG1DM (dot
  dashed), shows instead a cuspy profile down to the force resolution
  (red dashed vertical line as for DG1MR). The central density is
  about 10 times higher than in the runs with strong outflows. DG2 has
  a similar profile to DG1, while the lower resolution run or the run
  with diffuse SF have dense and cuspy DM profiles down to the force
  softening length.}
\label{rotcurve}
\end{figure}

\begin{figure}
\vskip 13.5cm
{\includegraphics{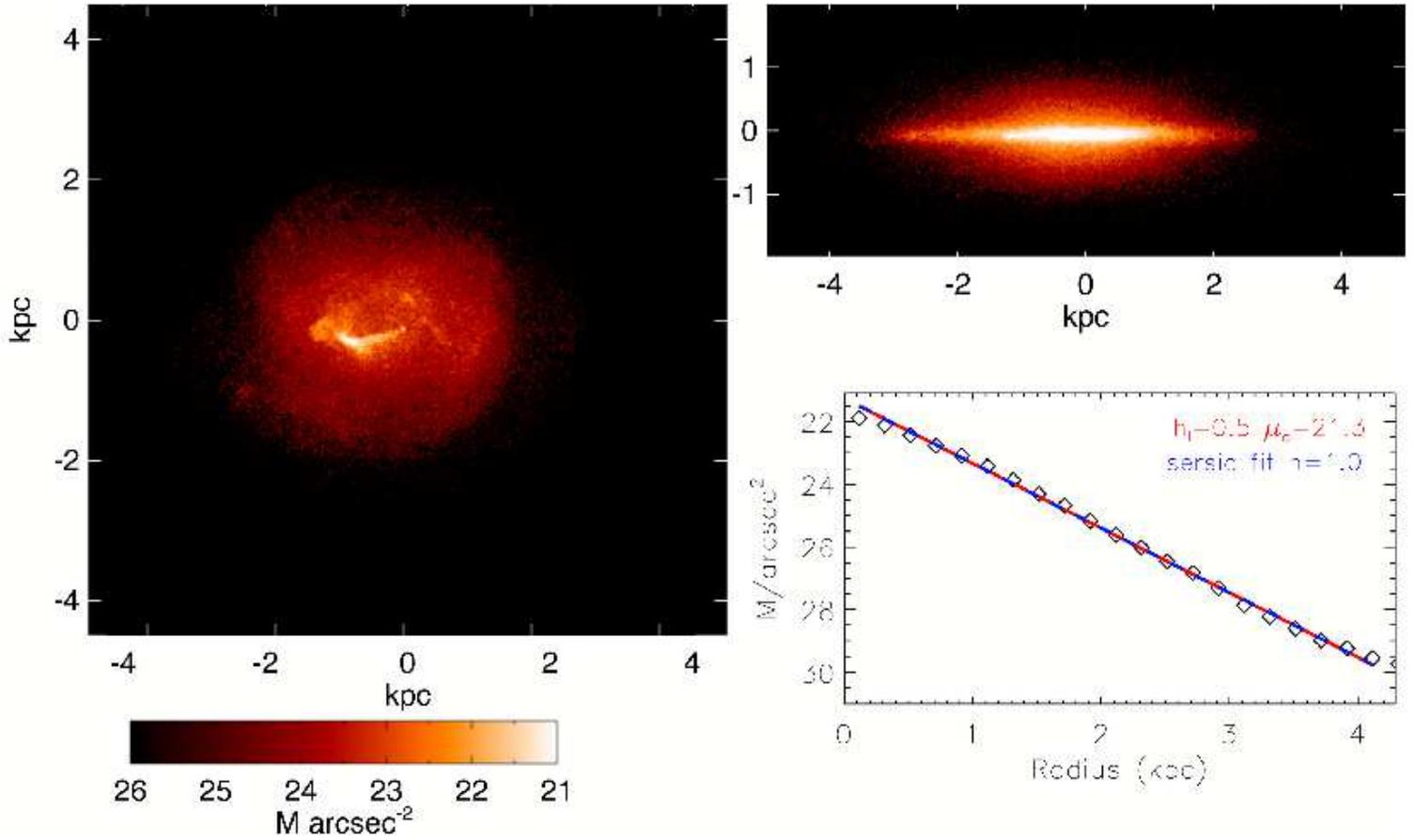}}
\caption[]{\small The optical properties of simulated galaxy
  DG2. Clockwise from left: The dust reddened $i$-band surface
  brightness map of galaxy DG2 seen face-on, edge-on, and the 1D
  radial surface brightness profile. The radial profile shows  that
  the simulated galaxy is ``bulgeless,'' with an almost perfect
  exponential profile with a scale length of 0.5 kpc. (the Sersic
  profile with n$=$1 is equivalent to an exponential profile).}
\label{DG2prof}
\end{figure}

\begin{table*}
\centering
\begin{tabular}{lrrrrrr}
\hline

Run & DM Part.    & Gas Part.   & Force           &  Redshift$_{LMM}$ & Stellar    & SF \\
    &Mass M$_{\odot}$ &Mass M$_{\odot}$ & Resol. (pc) & z~~~~~           & Mass M$_{\odot}$ & \\
\hline
\\
DG1     & 1.6e4  &  3.3e3 & 86  &1. & 4.8e8 & High Th. \\
DG1MR   & 3.7e4  &  7.8e3 & 116 &1. & 5.0e8  & High Th. \\
DG1LR   & 1e6    &  2.1e5 & 350 &1. & 4.8e8   & High Th. \\
DG1LT   & 3.7e4  &  7.8e3 & 116 &1. & 3.7e9   & Low Th.\\
DG1DM   & 2e4 &     -     & 86  &1. & -   &  - \\
DG2     & 1.6e4  &  3.3e3 & 86  &2. & 1.8e8  & High Th. \\
\hline
\end{tabular}
\caption[Parameters of the simulated dwarf galaxies]
{Summary of the main numerical parameters for each simulation described in \S3 and \S5. From left: DM and gas particle masses, force resolution, redshift of the last major merger, galaxy stellar mass at z$=$0 and the SF model adopted.The high resolution DG1 and DG2 galaxies  have a total virial mass of 3.5 (2.0) 10$^{10}$ M$_{\odot}$ respectively. They both have a $\lambda$ spin parameter$^{56}$ of  0.05. The highest resolution version of DG1 has   about 3.5 million particles within the virial radius. }
\label{parameters}
\end{table*}

\section{Inhomogeneous ISM, the removal of low angular momentum gas and the expansion of the central DM  distribution}

Several analytical and numerical papers have highlighted the necessity
of resolving a clumpy multi-phase ISM to achieve a realistic modeling
of energy deposition in the central regions of galaxies.  While using
a variety of arguments, these works suggest that only in a clumpy ISM
is it possible to a) transfer orbital energy from gas to the DM as
dense clumps sink through dynamical friction or resonant
coupling$^{8,16,21,22,59,60}$ and have b) efficient gas
outflows$^{15,54}$. In turn, these outflows 1) suppress the formation
of stellar bulges by removing negative or low angular momentum
gas$^{7,15,16,21}$ and 2) make the central DM expand by suddenly
reducing the total enclosed mass and reducing the DM binding
energy$^{21}$.  However, none of the above works has simultaneously
studied the formation of bulgeless galaxies and that of DM cores, even
if they both are crucial properties of small galaxies.

\begin{figure}
\vskip 13cm
{\includegraphics{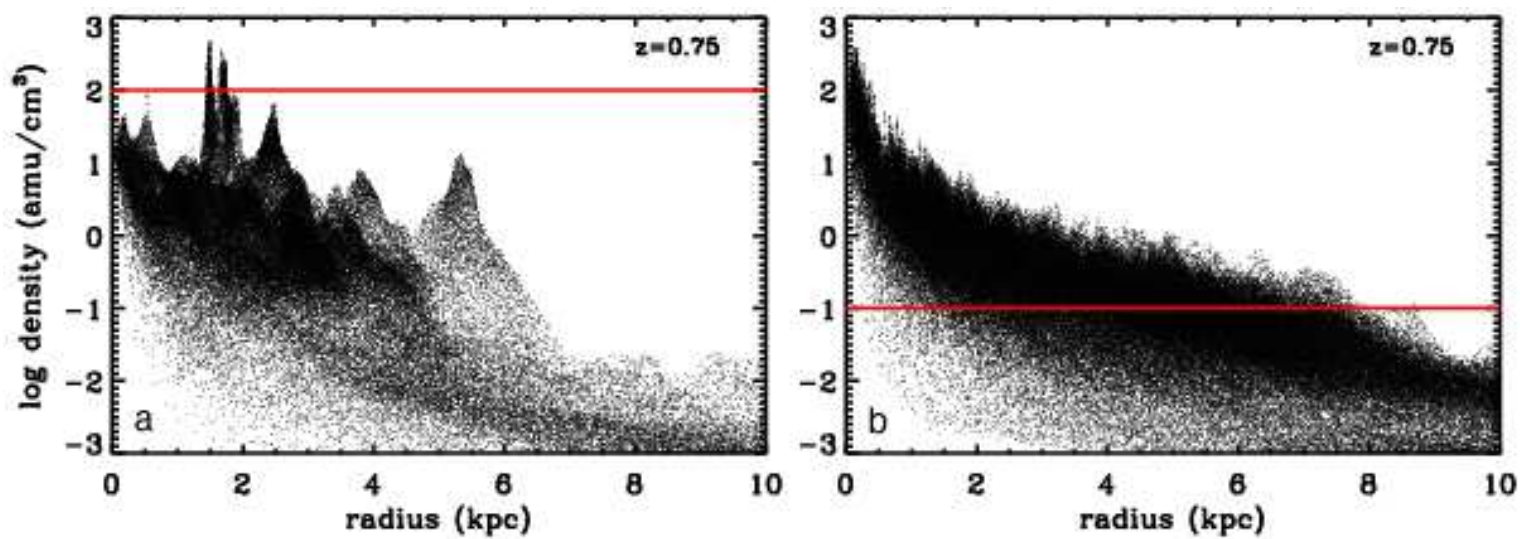}}
\caption[]{\small Properties of the gas distribution for different SF
  implementations. The local gas density measured around each SPH
  particle is plotted as a function of its radial distance from the
  galaxy center for runs DG1MR and DG1LT, which have identical force
  and mass resolution, but differ in the star formation density
  threshold.  Horizontal red lines mark the minimum gas density for SF
  in each run.  In both runs the SFR is $\propto$
  ${\rho_{gas}}^{1.5}$.  In the low threshold simulation, diffuse star
  formation in the inner regions continues unabated by feedback, as SN
  energy is more evenly distributed and is unable to originate major
  outflows.  Allowing star formation only in high density regions
  results in a complex, inhomogeneous ISM, even in the central regions
  and fast outflows that remove gas preferentially from the galaxy
  center.}
\label{dens_rad}
\end{figure}

In order to achieve a multi phase ISM numerical works agree that a
minimal spatial resolution of about 100pc is required, and that SF has
to be associated with dense regions with gas density ($\sim$ 100
amu/cm$^3$)$^{40,61}$.  Our simulations satisfy both conditions and
unify the many proposed models that focused on different aspects of
the problem to robustly demonstrate that energy transfer and
subsequent baryon removal are concurrent and effective to create
bulgeless galaxies with a shallow DM profile in a full cosmological
setting.

To illustrate the clumpiness of the ISM in our simulations, Figure~7
highlights the differences in the density distribution of the
interstellar medium between the simulations DG1MR and DG1LT, by
plotting the local gas density vs radius of each gas particle at a
representative epoch of z$=$0.75. These two runs have the same mass
and spatial resolution and adopt identical feedback schemes. They only
differ in the way regions where SF happens are selected.  In the
``high threshold'' runs (DG1,DG2 DG1MR and DG1LR) SF happens only in
regions above a high gas density threshold (100 amu/cm$^3$, the
horizontal red line in the left panel of Figure 7).  The density peaks
then correspond to isolated clumps of cold gas with masses and sizes
typical of SF regions.  The efficiency of SF, $\epsilon$SF, for these regions
must be increased from 0.05 (LT) to 0.1 (HT) in order to match the
observed normalization of SF density in local galaxies.  However, due
to the increased densities, at any given moment only a few regions are
actively forming stars.  These star forming regions get disrupted
after the first SNe go off and only a small fraction of gas has been
turned into stars. Feedback then  creates an ISM with cold filaments and
shells embedded in a warmer medium. This patchy distribution allows
the hot gas to leave the galaxy perpendicular to the disk plane at
velocities around 100 $\kms$.  Rather than developing a clumpy ISM as
in the ``high threshold'' case, SF in the ``low threshold'' scheme is
spatially diffuse (Figure~7).  This means that SN energy is more
evenly deposited onto the gas component, but less overall gas is
effected by SN feedback due to the low densities in the SF regions.
By monitoring where SN energy is deposited and where gas gets
substantially heated at high instantaneous rates, we verified that in
the ``high threshold'' case a larger mass of gas achieves temperatures
higher than the virial temperature (T$_{vir}$ $\sim$ 10$^5$ K) per
unit mass of stars formed than in the ``low threshold'' scheme.  Since
less mass is affected in the ``low threshold'' scenario, the outflows
are weak compared to the ``high threshold'' case.  By z$=$0 DG1LT has
formed ten times more stars, most of them in the central few kpcs,
causing strong adiabatic contraction of the DM component. Its light
profile is consistent with a B/D ratio of 0.3, typical of much more
massive galaxies and more concentrated than in real dwarfs.

We note that in the runs adopting the ``high threshold'' SF, feedback
produces winds that are comparable in strength to those happening in
real galaxies of similar mass. However, in our simulations the cold ISM is
still only moderately turbulent ($\sim$ 10 $\kms$ at z=0), consistent
with observations$^{62}$, and the galaxies match the observed stellar
and baryonic Tully Fisher relation$^{63}$, as the SF efficiency is
regulated to form an amount of stars similar to that of real dwarf
galaxies of similar rotation velocity.

\begin{table*}
\centering
\begin{tabular}{lcccccc}
\hline

Run & M$_i$  & $g-r$ & SFR  & R$_s$ $i$  & V$_{rot}$ & M$_{HI}$/L$_B$  \\
& &   &M$_{\odot}$/yr& kpc & km/s & $ $ \\
\hline
\\
DG1     & -16.8  & 0.52 & 0.01 & 0.9  & 56  & 1.2   \\
DG1MR   & -16.9  & 0.54 & 0.02 & 0.9  & 55  & 1.0   \\
DG1LR   & -18.7  & 0.33 & 0.22 & 0.9  & 62  & 0.64    \\
DG1LT   & -19.4  &  0.40 &  0.38 & 1.3  & 78  & 0.11   \\
DG2     & -15.9 &   0.46 & 0.02 & 0.5  & 54  & 2.8   \\
\hline
\end{tabular}
\caption[Summary of the observable properties of the simulated dwarf galaxies]
{Summary of the observable properties of the different dwarf runs. the SFR is in M$\odot$/yr, R$_s$ is the disc scale length,  V$_{rot}$ is measured using the HI velocity field as W$_{20}$/2}
\label{parameters}
\end{table*}

\begin{figure}
\vskip 10cm
{\includegraphics{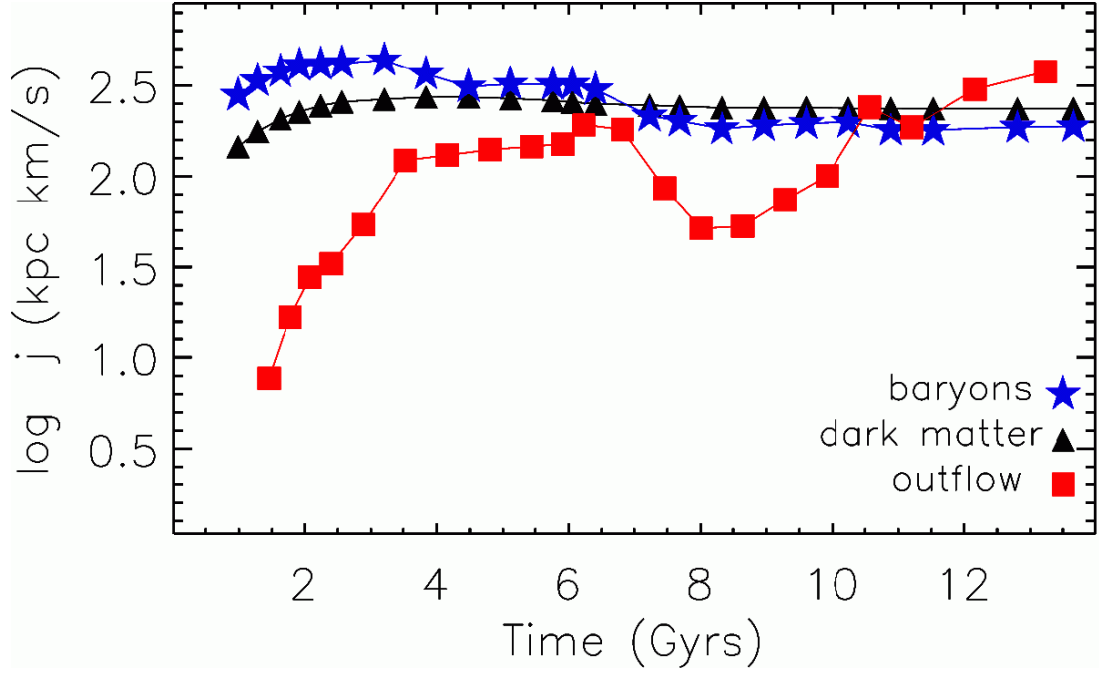}}
\caption[]{\small The angular momentum per unit mass of the DM and gas
  being accreted and of the gas being blown out as a function of
  time. While most of the matter accreted onto the galaxy halo has a
  fairly constant angular momentum, the gas being blown out at high
  redshift has a systematically lower angular momentum, by as much as
  an order of magnitude.  During the same epoch and until {\it z}$=$1
  the SFH of the galaxy is quite bursty. The stronger bursts are
  closely associated with epochs where the baryon distribution is
  clumpier and with decreases of the central DM density, supporting
  models$^{8,15}$ for the formation of the DM core and the formation of bulgeless galaxies.}
\label{fig:s3}
\end{figure}

Figure~8 illustrates an essential property of the simulations
presented in this Letter: outflows selectively remove low angular
momentum gas from high redshift galaxies. The mean angular momentum of
gas blown out of the virial radius of galaxy DG1 is plotted as a
function of time (red triangles) and shown to be as much as ten times
smaller than the mean angular momentum of all the baryons (blue) and
dark matter (black) accreted at the same time and then retained within
the virial radius of the galaxy at $z=0$. As the central region of a
galaxy is being assembled at z $>$ 1.5 its mass profile and angular
momentum content is then directly affected.  This scenario, that
maximizes energy transfer to the DM and gas outflows in the regions
that will become the center of the galaxy by z$=$0, is a natural
consequence of the hierarchical assembly of galaxies in the CDM model.

The results described in the Letter confirm that energy transfer and
subsequent gas removal in a clumpy ISM have the net effect of causing
the central DM distribution to expand, while at the same time limiting
the amount of baryons at the galaxy center. In Figure 5 we quantify
this flattening by showing that by the present time the DM central
profile in galaxies DG1 and DG2 is well approximated by a power law
with slope $\alpha$ in the -(0.5--0.7) range. These values of $\alpha$
are significantly flatter than in the collisionless control run and
are in agreement with those of observed shallow DM profiles in dwarf
galaxies$^{13,14,64}$.  This result also resolves the dichotomy
between CDM predictions of cuspy profiles based on DM--only
runs$^{46,57}$ and observations: neglecting the modeling of gas
outflows suppresses the removal of DM matter from the galaxy center
and makes the formation of shallow profiles impossible. The central
concentration of mass in DG1LT, in which outflows are suppressed
(Figure 5) and DM adiabatically contracts rather than expanding,
results in the very steep inner rise of the rotation curve, followed
by a steep decline. Such declining rotation curves are unrealistic in
small galaxies and have long plagued hydrodynamical galaxy formation
simulations. By looking at smaller, but well resolved halos formed in
the high resolution region of our simulations (but outside the virial
radius of the main galaxies) we verified that star forming galaxies
with rotational velocity V$_{rot}$ $\sim$ 30 $\kms$ also have shallow
DM profiles, while even smaller dark galaxies (where SF was completely
suppressed due to the gas heating by the cosmic UV field) have steep
central DM profiles. This result confirms that gas outflows caused by
SF feedback are the cause for the removal of low angular momentum gas
and the formation of DM cores.

\section{Supplementary Notes}

FG acknowledges support from HST GO-1125, NSF ITR grant PHY-0205413
(also supporting TQ), NSF grant AST-0607819 and NASA ATP
NNX08AG84G. CBB acknowledges the support of the UK's Science \&
Technology Facilities Council (ST/F002432/1). PJ was supported by
programs HST-AR-10678 and 10958 and by Spitzer Theory Grant 30183 from
the Jet Propulsion Laboratory. LM was supported by a grant of the
Swiss National Science Foundation. LM and CB thank the Kavli Institute
for Theoretical Physics at UC Santa Barbara for hospitality during
completion of the work.  We thank the computer resources and technical
support by TERAGRID, ARSC, NAS and the UW computing center, where the
simulations were run.

\section{Supplementary Movie Legend}

The movie shows the evolution of the gas density (in blue) in the
region where galaxy DG2 forms, from shortly after the Big Bang to the
present time.  The frame is 15 kpc per side.  Brighter colors
correspond to higher gas densities. The movie highlights the numerous
outflows that remove low angular momentum gas and the connection
between outflows with accretion and early merger events.
\end{document}